\documentclass[preprint,12pt,authoryear]{elsarticle}

\usepackage{amssymb}

\usepackage{url}
\usepackage{subfig}
\usepackage{multirow}
\usepackage{booktabs}

\usepackage{diagbox}
\usepackage{fnpct}
\usepackage{ulem}

\usepackage{amsmath}
\usepackage{amssymb}
\renewcommand{\vec}[1]{\boldsymbol{{#1}}}

\newcommand{\pten}{$P@\textrm{10}$ }
\newcommand{\pN}{$P@N$ }

\usepackage[prependcaption,textsize=scriptsize]{todonotes}
\definecolor{mycolor}{HTML}{FF6600}

\journal{Computer Speech and Language}

\begin{document}

\begin{frontmatter}

\title{Feature learning for efficient ASR-free keyword spotting in low-resource languages}

\author[label_su]{Ewald van der Westhuizen}
\author[label_su]{Herman Kamper}
\author[label_su]{Raghav Menon}
\author[label_plk,label_edn]{John Quinn}
\author[label_su]{Thomas Niesler\corref{cor1}}

\address[label_su]{Department of Electrical and Electronic Engineering, Stellenbosch University, South Africa}
\address[label_plk]{UN Global Pulse, Kampala, Uganda}
\ead{trn@sun.ac.za}

\cortext[cor1]{Corresponding author.}
\fntext[label_edn]{Part of this work was performed while John Quinn was a visitor at the University of Edinburgh, UK.}

\begin{abstract}
We consider feature learning for efficient keyword spotting that can be applied in severely under-resourced settings.
The objective is to support humanitarian relief programmes by the United Nations in parts of Africa in which almost no language resources are available.
For rapid development in such a language, we rely on a small, easily-compiled set of isolated keywords.
These keyword templates are applied to a large corpus of in-domain but untranscribed speech using dynamic time warping (DTW).
The resulting DTW alignment scores are used to train a convolutional neural network (CNN) which is orders of magnitude more computationally efficient and suitable for real-time application.
We optimise this neural network keyword spotter by identifying robust acoustic features in this almost zero-resource setting.
First, we consider incorporating information from well-resourced but unrelated languages using a multilingual bottleneck feature (BNF) extractor.
Next, we consider features extracted from an autoencoder (AE) trained on in-domain but untranscribed data.
Finally, we consider correspondence autoencoder (CAE) features which are fine-tuned on the small set of in-domain labelled data.
Experiments in South African English and Luganda, a low-resource language, show that BNF and CAE features achieve a 5\% relative performance improvement over baseline MFCCs.
However, using BNFs as input to the CAE results in a more than 27\% relative improvement over MFCCs in ROC area-under-the-curve (AUC) and more than twice as many top-10 retrievals.
We show that, using these features, the CNN-DTW keyword spotter performs almost as well as the DTW keyword spotter while outperforming a baseline CNN trained only on the keyword templates.
The CNN-DTW keyword spotter using BNF-derived CAE features represents an efficient approach with competitive performance suited to rapid deployment in a severely under-resourced scenario.
\end{abstract}


\end{frontmatter}

\section{Introduction}

Social media can be used effectively to monitor the views and concerns of a population for the purposes of informing relief and developmental programmes.
However, in many countries internet infrastructure is poorly developed, precluding this approach.
In such cases, community radio phone-in talk shows are used by citizens as a platform to voice issues of pressing importance.
In a project piloted by the United Nations (UN), radio browsing systems have been developed to monitor such radio shows in Uganda~\citep{Menon2017,Saeb2017}.
Currently, these systems are actively and successfully supporting relief and developmental programmes by the organisation.
However, the deployed radio browsing systems rely on automatic speech recognition (ASR), the development of which depends on the availability of transcribed speech data in the target languages.
This has proved to be a serious impediment when very quick intervention is required in a different location, since the development of such a corpus in a new language is always time-consuming, expensive and requires linguistic expertise.

The purpose of a keyword spotting system is to search audio speech data for a set of keywords.
The conventional approach is to perform ASR and generate lattices which are in turn searched for the presence or absence of these keywords~\citep{Larson12, mandal2014recent}.
In resource-constrained settings, the training data required to develop ASR may not be available.
In these circumstances keyword spotting approaches that can be developed without substantial labelled data become attractive~\citep{chen2014small, SainathPara2015, audhkhasi+etal_icassp17, tang2018deep}.
We will refer to such keyword spotting systems that do not rely on ASR as \textit{ASR-free}.

One approach to ASR-free keyword spotting is to extend query-by-example search (QbE), where the search query is provided as audio rather than as text.
QbE can be achieved by using dynamic time warping (DTW) to perform a direct match between a search query and utterances in the search collection~\citep{park+glass_taslp08,Hazen2009, Zhang2009,jansen+vandurme_interspeech12}.
For keyword spotting, this approach requires one or more labelled spoken keyword instances as templates.
These templates are used as queries for DTW-based QbE.
Since the class of each template is known, the individual per-exemplar QbE results can be aggregated to determine whether a certain keyword occurs in a particular utterance.
The advantage of this keyword spotting approach is that only a small set of labelled keywords is required and not a large transcribed corpus as used for ASR-based keyword spotting.
A key disadvantage is that DTW is computationally very expensive and usually not feasible for large-scale continuous application~\citep{Menon_Inter2018}.
Furthermore,  for DTW the choice of input features has a greater impact on performance than, for example, for ASR, since in the latter case acoustic models can learn to adapt to different representations~\citep{Menon_SLTU2018}.

In this paper, we consider a low-resource setting in which we have at our disposal a small, easily acquired and independently compiled set of isolated keywords in the language of interest, a larger body of in-domain but untranscribed speech also in the target language, and a collection of out-of-domain labelled speech data from other well-resourced languages.
The isolated keywords are collected separately and are not drawn from the radio speech domain to which our keyword spotters are applied.
Acquiring these resources in a new and under-developed language is much easier than compiling a corpus of transcribed speech because no annotation in the new language is required.
We will show that these different resources can be complimentary in enabling robust feature learning and improved efficiency for low-resource ASR-free keyword spotting.
To do this, we perform experiments on radio broadcast data from two languages: South African English, a fairly well-resourced language on which we perform all development experiments, and Luganda, a low-resource Ugandan language of current interest for humanitarian relief efforts.

We achieve computationally efficient ASR-free keyword spotting by training a convolutional neural network (CNN) to emulate DTW-based scoring.
To achieve this, we use the small corpus of isolated keywords to perform DTW template matching on the larger corpus of untranscribed data in the target language and domain.
The resulting DTW scores are used to train a CNN, which can be applied to new data instead of DTW.
In this way we take advantage of CNN-based searching, which is computationally efficient since it does not require alignment, to perform DTW-based matching, which requires a minimum of labelled data.
We first proposed this approach, which we will refer to as CNN-DTW keyword spotting, in~\citep{Menon_Inter2018} and later extended it in~\citep{Menon_SLTU2018} and in \citep{menon+etal_interspeech19}.

As already pointed out, DTW is more sensitive to the features used to represent the speech signal than ASR.
To address this and allow the learning of improved frame-level acoustic features, we build on recent work in ``zero-resource'' speech processing, where the goal is to learn robust feature representations without access to any labelled speech data~\citep{versteegh2016zero,dunbar2017zero,dunbar+etal_interspeech19}.
Various different features and learning approaches have been considered ranging from conventional speech features~\citep{carlin2011rapid,vavrek2012tuke,lopez2016finding}, to posteriorgrams from probabilistic mixture models~\citep{Zhang2009,jansen2012jhu,heck+sakti,heck+etal_ieice18}, to latent representations computed by neural networks~\citep{badino+etal_interspeech15,renshaw+etal_interspeech15,triamese,riad2018sampling,eloff+etal_interspeech19}.
Among these, multilingual bottleneck feature (BNF) extractors, trained on well-resourced but out-of-domain languages, have been found by several authors to improve on the performance of MFCCs and other representations~\citep{vesely2012language,vu2012investigation,sthomas2012,cui2015multilingual, alumae2016improved,chen2017multilingual, yuan2017extracting,Hermann2018,hermann+etal_submitted18}.

We will further improve on the performance of BNFs by fine-tuning on the small set of isolated keywords we have available in the target low-resource language.
We achieve this by employing a correspondence autoencoder (CAE), a model also originally developed for the zero-resource setting~\citep{kamper2015unsupervised,renshaw+etal_interspeech15}, to learn a mapping between all possible combinations of alternative utterances of the same keyword type in our small dataset.
This strategy trains the CAE to disregard aspects of the acoustics not common to the presented keyword utterances, such as speaker, gender and channel, while capturing aspects that they do have in common, such as word identity.
Using these CAE$_{\rm BNF}$ features, the CNN-DTW keyword spotter performs almost as well as the DTW-based system used to supervise it during training, but is three orders of magnitude faster at runtime.
This approach was inspired by ideas established in \citep{Hermann2018} and~\citep{hermann+etal_submitted18}, where a CAE trained on BNFs using a large set of in-corpus word pairs outperformed other methods in intrinsic evaluations.
In that case, however, the improvement was shown not to hold consistently in the completely unsupervised case where automatically discovered word segments were used as supervision for the CAE.

While this paper builds on our previous work, it includes a number of important new developments.
\begin{itemize}
\item While previous work has focused on English as a convenient development language due to its better resources, we now also apply the CNN-DTW approach to a truly low-resource language, Luganda.
\item We apply CAE, use the CAE$_{\rm BNF}$ features as inputs to the CNN-DTW model and show that this combination yields performance comparable to the DTW system on which it is based.
\item We benchmark the CNN-DTW against the much more computationally expensive DTW classifier as well as a straightforward end-to-end CNN classifier and show that the CNN-DTW is computationally more efficient by several orders of magnitude.
\end{itemize}
These additional results allow us to conclude that the use of supervised features trained on out-of-domain labelled data from well-resourced languages are complementary to features learned by fine-tuning on a small set of in-domain data, and that keyword spotting efficiency can be improved by training the CNN-DTW on targets automatically obtained on a larger corpus of unlabelled in-domain speech.
 \section{Radio browsing}

We start by presenting the wider context for this work by describing the radio browsing application in which our models will be used and the data on which the models are trained.

\subsection{Radio browsing system}

\begin{figure}[!ht]
\centering
  \includegraphics[width=0.96\textwidth]{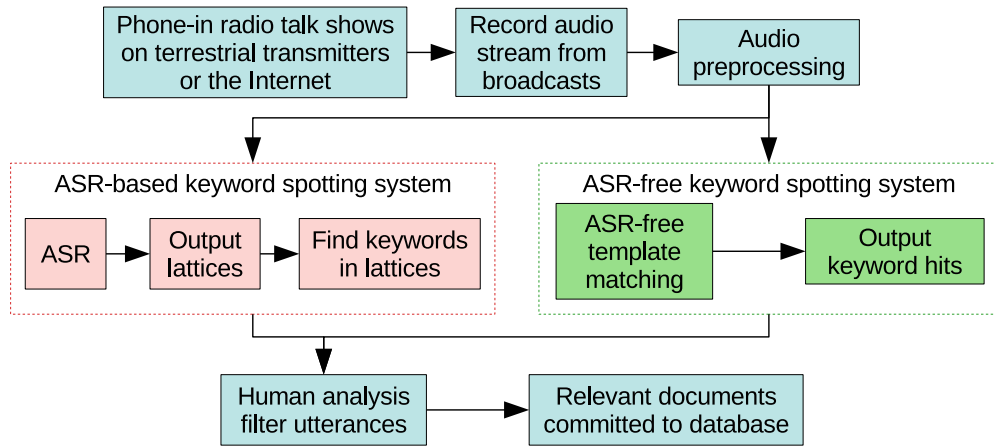}
  \caption{The workflow of the UN Global Pulse radio browsing system. The ASR-based and ASR-free alternatives to keyword spotting are shown in red and green, respectively.}
  \label{fig:radio_browsing_system}
\end{figure}

Local radio stations, specifically in Uganda, host phone-in talk shows where listeners can call in and discuss issues of current interest.
These have in the past included topics such as flooding, malaria, cholera, violence against women, adolescent pregnancy, and price fluctuations of local goods.
The radio shows are typically broadcast in the local language of the region.
In our radio browsing configuration, illustrated at a high level in Figure~\ref{fig:radio_browsing_system}, broadcasts are recorded using RTL2832U-based DVB-T receivers, Raspberry Pi computer platforms and the GNU Radio open source software tools\footnote{\url{https://www.gnuradio.org/}}.
This system allows multiple radio stations to be captured on each Raspberry Pi.

Recordings are processed using a keyword spotting system, which identifies utterances that may contain specific words of interest.
This set of keywords is determined and refined by the humanitarian agency, since the goal is to monitor and address specific societal concerns.
Currently, keyword spotting is performed using an ASR system, which converts the input speech to lattices which are subsequently searched for the keywords.
This approach is illustrated on the left side of Figure~\ref{fig:radio_browsing_system}.
Human analysts filter the utterances identified in this way and add useful detections to a searchable database that supports decision making regarding humanitarian interventions\footnote{{Examples of identified radio broadcasts can be found at \mbox{\url{https://radio.unglobalpulse.net/uganda}.}}}.

To develop the required ASR system, a hand-annotated corpus of transcribed speech in the target language was required~\citep{Saeb2017}.
However, the collection of even a small, fully-transcribed corpus has proven difficult or impossible for many of the local languages spoken in the region, due to constraints in time, funding and language expertise.
This has motivated us to consider ASR-free keyword spotting, as illustrated on the right side of Figure~\ref{fig:radio_browsing_system}.
In this approach, ASR is replaced by a system which directly identifies whether a keyword is present in an utterance or not, without converting the incoming speech to intermediate lattices or text.

\subsection{Data}
\label{sec:data}

Our ASR-free keyword spotting experiments are performed in two languages: South African English and Luganda.
English is used because the availability of annotated speech allowed more thorough development experiments to be performed.
In contrast, keyword spotting in Luganda represents a practical application of a radio-browsing system in a truly low-resource setting.

\subsubsection{English and Luganda in-domain corpora}
\label{sec:radio_corpora}

For English, we use a corpus of South African Broadcast News (SABN) which consists of 23 hours of speech compiled from news bulletins broadcast between 1996 and 2006 by one of {South Africa's} main radio news channels~\citep{Kamper2015}.
The corpus contains a mix of newsreader speech, interviews and crossings to reporters.
About 80\% of the speakers can be considered native English speakers.

The Luganda data was collected from radio broadcasts in Kampala, Uganda.
It was noted subjectively that, in comparison with the SABN data, these radio recordings often seemed to contain more noise.
The quality was particularly low for some of the phone-in speech, which suffered from distortions resulting from both FM transmission and mobile phone data compression.
This however represents the practical setting in which the keyword spotting systems will have to perform.

For our experiments, the English and the Luganda data are divided into training, development and test partitions, as shown in Table~\ref{tab:radio_data}.
For both languages, the respective training sets are used {exclusively} to fulfil the role of the in-domain untranscribed data that will be used to obtain DTW targets with which to train the CNN and also to pretrain the autoencoder feature extractor.
Keyword spotting is then performed on the test sets, which fulfil the role of search data, using their transcriptions to evaluate performance.
Hyperparameter optimisation is performed on the English development data, after which the approach is applied without any further tuning to both the English and the Luganda test sets.

\begin{table}[!t]
\small
\centering
\caption{The South African English and Luganda datasets, indicating number of utterances and durations for the training, development and test partitions.}
\label{tab:radio_data}
\renewcommand{\arraystretch}{1.1}
\begin{tabular*}{0.96\textwidth}{@{\extracolsep{\fill}} lcccc @{}}
\toprule
\multirow{2}{*}{\textbf{Set}} & \multicolumn{2}{c}{\textbf{English}} & \multicolumn{2}{c}{\textbf{Luganda}} \\
\cmidrule(lr){2-3} \cmidrule(l){4-5} & \textbf{Utterances} & \textbf{Duration} & \textbf{Utterances} & \textbf{Duration}      \\ \midrule
Training     & \leavevmode\hphantom{0}5\,231 & \leavevmode\hphantom{0}7.94h &     6\,052      &         5.57h          \\
Development  & \leavevmode\hphantom{0}2\,740 & \leavevmode\hphantom{0}5.37h &     1\,786      &         2.04h          \\
Test         & \leavevmode\hphantom{0}5\,005 &                       10.33h &     1\,420      &         1.99h          \\
\midrule
Total        &                       12\,976 &                       23.64h &     9\,258      &         9.06h          \\ \bottomrule
\end{tabular*}
\end{table}

\subsubsection{English and Luganda keyword corpora}
\label{sec:keyword_data}

For English, a small independent corpus of 40 isolated keywords, each uttered at least once by 24 South African speakers (12 male, 12 female) was compiled.
The resulting set of 1160 isolated keyword utterances, amounting to around 20 minutes of speech, represents the only labelled in-domain English data used to train the keyword spotting system.
There is no speaker overlap with the English SABN dataset (Section~\ref{sec:radio_corpora}).

Similarly, the Luganda keyword corpus is an independent collection of 18 keyword types uttered by several male and female speakers.
The recording conditions are highly variable and often include substantial background noise.
Approximately 32 utterances per keyword type were retained after performing quality control on the recordings.
The resulting set of 603 isolated keyword utterances, amounting to approximately 13 minutes of speech, represents the only labelled Luganda in-domain data used to train the keyword spotting system.
There is no speaker overlap with the Luganda radio dataset (Section~\ref{sec:radio_corpora}).

A summary of the composition of the keyword corpora is given in Table~\ref{tab:kw_dataset}.
The mismatch between the keyword corpora in  Table~\ref{tab:kw_dataset} and the search datasets in Table~\ref{tab:radio_data} for both languages is intentional as it reflects the operational setting of our low-resource radio browsing systems, where the small labelled training set is compiled from a different set of speakers and under different recording conditions than the search data.

\begin{table}
\small
\renewcommand{\arraystretch}{1.1}
\caption{The isolated keyword datasets for English and Luganda, indicating the number of keyword types, speakers and utterances.}\label{tab:kw_dataset}
\begin{tabular*}{\textwidth}{@{\extracolsep{\fill}} l c c c @{}}
\toprule
\textbf{Language} &  \textbf{Keywords} & \textbf{Speakers} & \textbf{Utterances} \tabularnewline
\midrule
English & 40   & 24 &           1\,160 \tabularnewline
Luganda & 18   & 16 & \phantom{0}\,603 \tabularnewline
\bottomrule
\end{tabular*}
\end{table}

\subsubsection{Multilingual bottleneck feature extractor training data}
\label{sec:bnf_data}

A bottleneck feature (BNF) extractor, described in more detail in Section~\ref{sec:bnf}, is trained on transcribed speech in several well-resourced languages that form part of the GlobalPhone corpus \citep{schultz+etal_icassp13}.
We use the BNF extractor developed by \cite{Hermann2018}, which is trained on a combined 198 hours of data from ten well-resourced languages: Bulgarian, Czech, French, German, Korean, Polish, Portuguese, Russian, Thai, and Vietnamese.
Note that this set does not contain either of the languages in which we perform keyword spotting.
 \section{ASR-free keyword spotting approaches}
\label{sec:kws_approaches}

We will consider three approaches to ASR-free keyword spotting that can be applied in our low-resource setting.
All three can be implemented given only a small set of isolated keyword audio templates and a larger corpus of unannotated speech in the target language.

\subsection{Convolutional neural network (CNN) keyword classifier}
\label{sec:cnn_classifier}

\begin{figure}[h!]
\includegraphics[width=\textwidth]{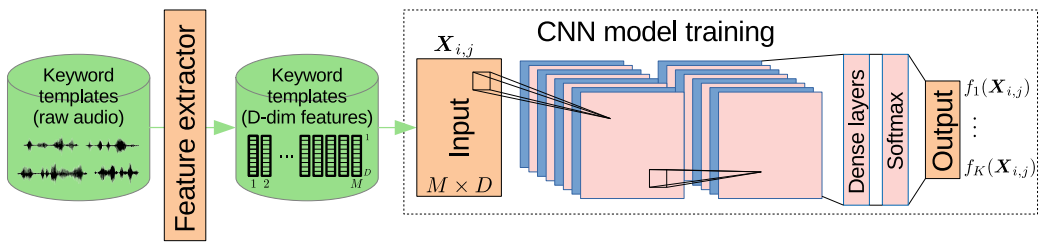}
\caption{The structure of the convolutional neural network classifier. To maintain clarity, not all network layers are shown.}
\label{fig:cnn_classifier}
\end{figure}

When a large number of labelled spoken keywords are available, an end-to-end keyword spotter can be trained to directly classify whether a keyword is present in a search utterance or not~\citep{SainathPara2015, palaz+etal_interspeech16, Kamper2016}.
Although we have only a limited number of such labelled keywords (Section~\ref{sec:keyword_data}), we nevertheless attempt to train a convolutional neural network (CNN) classifier in this way.
Assume the $D$-dimensional frame-level acoustic feature vector sequence for the
$j^\textrm{th}$ template of the $i^\textrm{th}$ keyword type is represented by \mbox{$\vec{X}_{i,j} = \vec{x}_1, \vec{x}_2, \ldots, \vec{x}_M$} where $\vec{x}_m \in \mathbb{R}^D$ for $1 \leq m \leq M$ and $M$ corresponds to the number of feature vectors in $\vec{X}_{i,j}$.
Our classifier accepts an $M \times D$ dimensional input ($M=60, D=39$) into a number of convolutional and max pooling layers.
These are followed by three dense layers terminated by a softmax layer with $K$ outputs, one for each keyword type.
This is illustrated in Figure~\ref{fig:cnn_classifier} while the precise network structure and implementation are presented in Section~\ref{sec:exp_setup_kws}.

Let the $k^\textrm{th}$ softmax output in response to an input keyword template $\vec{X}_{i,j}$ be $f_k(\vec{X}_{i,j})$.
The CNN classifier is trained using the set of separately recorded isolated keywords in a supervised fashion using the categorical cross entropy loss, which for the single training example $\vec{X}_{i,j}$ is given by:

\begin{equation}
    \ell = - \sum_{k = 1}^K J_k \, \log f_k(\vec{X}_{i,j}) = - \log f_i(\vec{X}_{i,j})
    \label{eq:multiclass}
\end{equation}

where $K$ is the number of keyword types and $J_k \in \{ 0, 1 \}$ is an indicator for whether $\vec{X}_{i,j}$ is an instance of keyword type $k$.
The $K$-dimensional softmax output of the CNN
$\vec{f}(\vec{X}_{i,j}) \in [0, 1]^K$ can therefore be interpreted as the estimated joint probability distribution over the set of $K$ keyword types for the particular input $\vec{X}_{i,j}$.

At test time, a sliding window of $M$ consecutive feature vectors is extracted from a test utterance $\vec{Y} = \vec{y}_1, \vec{y}_2, \ldots, \vec{y}_N$ and presented to the CNN.
This results in a sequence of $K$-dimensional classification results for the test utterance.
The maximum among these sequential scores for each of the $K$ dimensions is taken to be the score indicating whether the particular keyword is present in the test utterance or not.
Finally, keyword presence in the utterance $\vec{Y}$ is determined by applying a threshold to each of these $K$ maxima.

\subsection{Dynamic time warping (DTW) keyword classifier}
\label{sec:dtw_classifier}
When only a few isolated spoken keywords are available, dynamic time warping (DTW) is an appropriate detection technique since it can be applied with as little as a single audio template~\citep{Bagnall2017}.
DTW aligns two sequences of feature vectors by warping their time axes to achieve the best match.
The alignment cost associated with this match can then be used as a measure  of similarity between the sequences.

Formally, assume two frame-level acoustic feature vector sequences \mbox{$\vec{X} = \vec{x}_1, \vec{x}_2, \ldots, \vec{x}_M$} and \mbox{$\vec{Y} = \vec{y}_1, \vec{y}_2, \ldots, \vec{y}_N$} where $\vec{x}_m \in \mathbb{R}^D$ and $\vec{y}_n \in \mathbb{R}^D$ for $1 \leq m \leq M$ and $1 \leq n \leq N$.
The computation of the DTW alignment cost will be indicated by $J = \textrm{DTW}(\vec{X},\vec{Y})$.
In our implementation we use the cosine similarity for frame-wise comparisons.
This similarity score, which lies in the interval [$-1, 1$], is then normalised by adding an offset of one and scaling by a factor $0.5$.
This results in a normalised similarity score that has a value of one when $\vec{X}$ and $\vec{Y}$ match exactly and a value approaching zero when they are very dissimilar.

To determine whether a keyword is present in an utterance, a na\"{i}ve method is to slide each template $\vec{X}_{i,j}$ over the search utterance $\vec{Y}$ and compute the DTW similarity $J_{i,j,q}$ within each window of overlap as follows:

\begin{equation*}
J_{i,j,q} = \textrm{DTW}\big(\vec{X}_{i,j},\vec{Y}(q,q+M-1)\big)
\end{equation*}
where $\vec{X}_{i,j}$ represents the $j^\textrm{th}$ template of the $i^\textrm{th}$ keyword type and $M$ corresponds to the number of feature vectors in $\vec{X}_{i,j}$ while $\vec{Y}(q,r)$ denotes the subsequence $\vec{y}_q, \vec{y}_{q+1}, \ldots, \vec{y}_r$ of the search utterance $\vec{Y}$.
More advanced approaches to finding matching subsequences using DTW have been proposed, for example by \cite{park+glass_taslp08} and by \cite{jansen+vandurme_interspeech12}, but we will restrict ourselves to this simpler implementation.
We let the window of overlap correspond to the length of the template and use a skip of 3 frames.

Since we have multiple templates for each keyword type, the final score indicating whether keyword type $i$ occurs in utterance $\vec{Y}$ is taken as the highest similarity score over all windows and over all templates of type $i$.

\begin{equation*}
    \hat{J}_i = \max_{\forall j,q} J_{i,j,q}
\end{equation*}

By applying an appropriate threshold to this score, a decision can be taken regarding the presence or absence of the keyword in unlabelled speech.
Although useful in a low-resource setting, a major disadvantage of the DTW-based search described above is that the repeated alignment between keywords and search utterances can be prohibitively computationally expensive.

\subsection{CNN-DTW keyword classifier}
\label{sec:cnn_dtw_classifier}

\begin{figure}[h!]
\includegraphics[width=\textwidth]{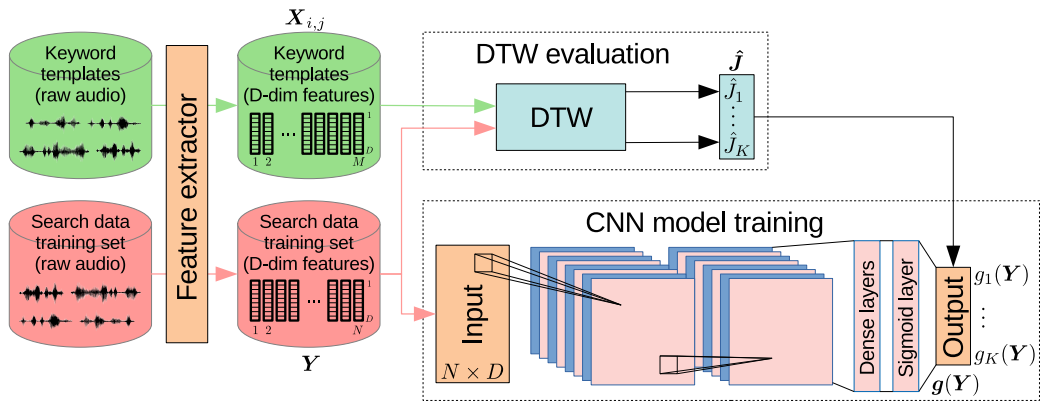}
\caption{A high-level diagram showing the concept of CNN-DTW training. To maintain clarity, not all network layers are shown.}\label{fig:cnn_dtw_concept_diagram}
\end{figure}

CNNs require large amounts of data for training, but are computationally efficient to apply.
DTW-based keyword spotting, on the other hand, can be applied with only a few keyword exemplars, but is computationally costly.
Our proposal is to employ DTW during training to address the challenge of data scarcity while taking advantage of the speed benefits of CNNs at runtime.
We achieve this by using DTW to calculate similarity scores between our small set of isolated keywords (Section~\ref{sec:keyword_data}) and the much larger but untranscribed in-domain dataset (Section~\ref{sec:radio_corpora}) and then using this set of similarity scores as targets to train a CNN.
An overview of this strategy is shown in Figure~\ref{fig:cnn_dtw_concept_diagram} while the implementation details and specific network architecture are discussed in Section~\ref{sec:exp_setup_kws}.

In the upper half of the figure, the $j^\textrm{th}$ template of the $i^\textrm{th}$ keyword type $\vec{X}_{i,j}$ is aligned with an utterance $\vec{Y}$ from the untranscribed data using DTW.
Subsequently, the highest similarity $\hat{J}_i$ among all the templates for the $i^\textrm{th}$ keyword type is determined.
This procedure is repeated for all keyword types.
The result is a vector of scores $\vec{\hat{J}} = [\hat{J}_1, \hat{J}_2, \ldots \hat{J}_K]$ for each utterance $\vec{Y}$ in the untranscribed corpus.
Each dimension of this vector gives an indication of whether the keyword of the corresponding type is present in the particular utterance.
Up to this point, this procedure is equivalent to the DTW-based keyword spotting approach described in Section~\ref{sec:dtw_classifier}.
However, instead of using $\vec{\hat{J}}$ to make a classification decision for each keyword type, we use these scores as targets for supervised training of a CNN.

Formally, given an unlabelled utterance $\vec{Y}$, the CNN-DTW model is trained to predict the associated vector of DTW scores $\vec{\hat{J}}$.
Our CNN consists of a number of convolutional layers, a global temporal max pooling layer, and a number of fully connected layers.
The global max pooling layer takes the maximum of the activations over the time dimension and therefore gives a fixed-dimensional output independent of the length of the input sequence.
The intuition is that this would extract the dominant features that are necessary for detecting the presence of a keyword in the utterance.
The final layer uses sigmoid activations to produce the model output $\vec{g}(\vec{Y}) = [g_1(\vec{Y}), g_2(\vec{Y}),\ldots,g_K(\vec{Y})]$ where $g_k(\vec{Y})$ is the output of the $k^\textrm{th}$ sigmoid in the final layer.
Note that, in contrast to the CNN classifier described in
Section~\ref{sec:cnn_classifier}, here the final layer is not a softmax.
Since more than one keyword may be present in the utterance $\vec{Y}$, the $K$ elements of $\vec{g}(\vec{Y})$ do not sum to one.
However, both the output of the network and the target DTW similarity scores are normalised to lie in the interval $[0, 1]$ and therefore resemble probabilities.
This allows us to train the CNN using the summed cross-entropy loss, which for utterance $\vec{Y}$ is given by:
\begin{equation}
    \ell = - \sum_{k = 1}^K \bigg\{ \hat{J}_k \, \log g_k(\vec{Y}) + (1 - \hat{J}_k) \log \big(1 - g_k(\vec{Y})\big) \bigg\}
    \label{eq:summed_binary}
\end{equation}

Note that because, in contrast to the loss in Equation~\eqref{eq:multiclass}, $\vec{g}(\vec{Y})$ is not a distribution over the keyword types, our CNN model can be considered a collection of $K$ binary classifiers, one for each keyword, with shared input layers.
The overall CNN-DTW model can then be trained using only a small set of labelled keywords and a large corpus of untranscribed speech.
 \section{Features and feature extractors}
\label{sec:features}

\begin{figure}
\centering
\includegraphics[width=0.65\textwidth]{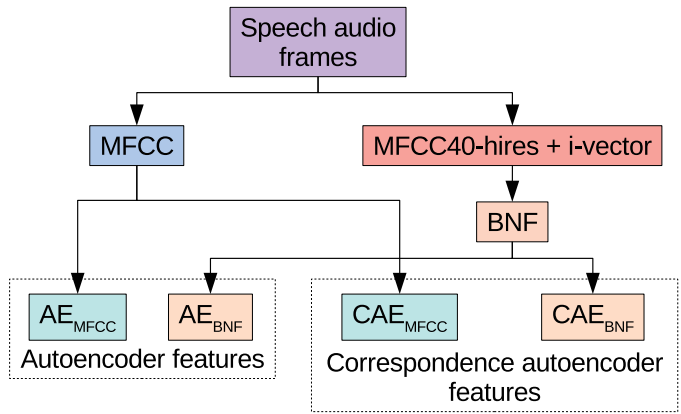}
\caption{The relationship between the feature types used for experimentation. Incoming arrows indicate input features used during training. For instance, for the CAE$_{\textrm{BNF}}$ features, BNFs are used to parameterise input speech for training a correspondence autoencoder.
}\label{fig:featureflow}
\end{figure}

We investigate the effect of different input features types for the keyword spotting approaches described in the previous section.
Mel-frequency cepstral coefficients (MFCCs) are used as baseline features and also serve as a basis for training the neural network-based feature extractors.
While transcribed in-domain data is difficult, time-consuming and expensive to compile, untranscribed in-domain speech data is much easier to obtain in substantial quantities.
We investigate the use of autoencoders (AEs) and correspondence autoencoders (CAEs) as a means of taking advantage of such untranscribed data.
The AE is trained on the untranscribed in-domain speech data, while the CAE is subsequently fine-tuned on the small set of labelled keyword exemplars in the target language.
In addition, although large amounts of transcribed in-domain speech data may not be available, large annotated speech resources do exist for several well-resourced languages.
These datasets can be used to train multilingual bottleneck feature (BNF) extractors.
Figure~\ref{fig:featureflow} summarises the different feature types we consider and the source features from which they were extracted.

\subsection{Mel-frequency cepstral features}
\label{sec:mfccs}

Mel-frequency cepstral coefficients (MFCCs) serve as baseline features.
From each 25ms speech frame we extract 13 MFCCs at a frame rate of 100Hz.
Velocity and acceleration coefficients are appended resulting in the classic 39-dimensional MFCC feature vectors.
These 39-dimensional MFCCs are used as input features for the AE$_{\rm MFCC}$ and CAE$_{\rm MFCC}$ neural networks described below.
For training of the multilingual BNFs, high resolution (40-dimensional) MFCCs are computed.

\subsection{Multilingual bottleneck features}
\label{sec:bnf}

\begin{figure}[!h]
\centering
\includegraphics[width=0.5\textwidth]{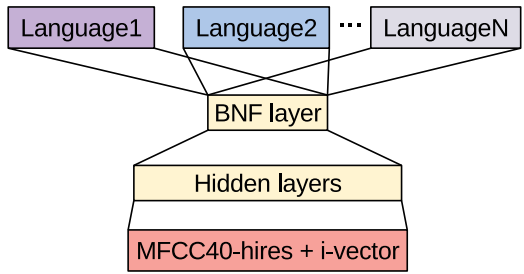}
\caption{Architecture of the bottleneck feature (BNF) extractor during training.}\label{fig:bnf_arch2}
\end{figure}

Multilingual bottleneck feature (BNF) extractors trained on a set of well-resourced languages have been shown to perform well in a number of studies and can be applied directly in an almost zero-resource setting~\citep{vesely2012language,vu2012investigation, cui2015multilingual, alumae2016improved,sthomas2012,chen2017multilingual, yuan2017extracting,Hermann2018,hermann+etal_submitted18}.
BNFs are obtained by joint training of a deep neural network on transcribed data from multiple languages, as shown in Figure~\ref{fig:bnf_arch2}.

The hidden layers of the network are shared among all languages.
The output layer has phone or HMM state labels as targets and may either also be shared or be separate for each language.
The layer directly preceding the output layer often has a lower dimensionality than the preceding layers, giving rise to the term ``bottleneck.''
The idea is that during training this layer is forced to capture aspects of the speech signal that are common to all the languages.
Features from this layer can then be used in a downstream task, even for an unseen language~\citep{Hermann2018,hermann+etal_submitted18}.

Different neural network architectures can be used to obtain BNFs.
We used the trained models that are described in~\citep{Hermann2018}.
This architecture uses the time-delay neural network (TDNN) structure and is trained on ten languages from the GlobalPhone corpus.
Implementation-specific details are given in Section~\ref{sec:bnf_setup}.
Note that the ``MFCC40-hires + i-vector'' features used to train the BNF extractor should not be confused with the baseline 39-dimensional MFCC features from the previous section.

\subsection{Autoencoder features}
\label{sec:ae_features}

An autoencoder (AE) is a feedforward neural network trained to reconstruct its input at its output.
A single-layer AE consists of an input layer, a hidden layer and an output layer.
In this case the AE takes input $\vec{x} \in \mathbb{R}^{D}$ and maps it to a hidden representation $\vec{h}=\sigma(\mathbf{W}^{(0)}\vec{x}+\vec{b}^{(0)})$, with $\sigma$ denoting a non-linear activation.
The output of the AE is obtained by decoding the hidden representation: $\hat{\vec{x}}=\mathbf{W}^{(1)}\vec{h}+\vec{b}^{(1)}$.
The network is trained to reconstruct the input using the loss $||\vec{x} - \hat{\vec{x}}||^2$.

\begin{figure}[!t]
\centering
\includegraphics[width=0.85\textwidth]{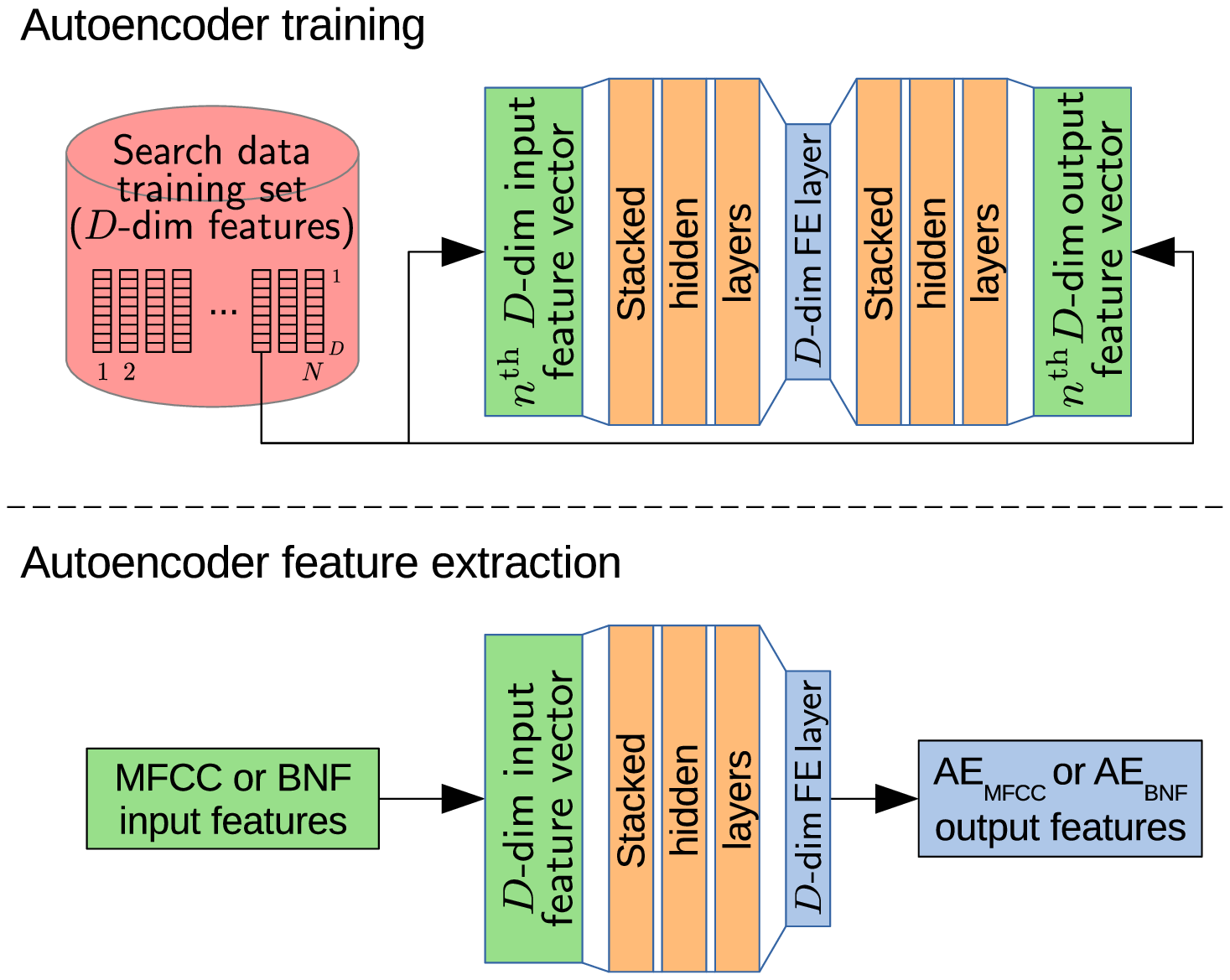}
\caption{
A simplified block diagram of the stacked autoencoder (AE) architecture during training (top) and feature extraction (bottom).
The feature extraction (FE) layer is shown in blue.
\textit{Top:} The network is trained on either MFCC features or BNFs by presenting the same feature vector at both the input and output layers.
A dimension of $D=39$ is used for the input, output and also the feature extraction layers.
\textit{Bottom:}
The trained AE is subsequently used as a feature extractor.
Furthermore, the stacked hidden layers are used to initialise the correspondence autoencoder (CAE) in Figure~\ref{fig:cae_train_extract}.
}\label{fig:ae_arch}
\end{figure}

A stacked AE~\citep{gehring2013extracting} is obtained by stacking several AE{s}, each AE-layer taking as input the encoding from the previous layer.
The stacked network is trained one layer at a time, each layer minimizing the loss of its output with respect to its input.
A number of studies have shown that hidden representations from an intermediate layer in such a stacked AE are useful as features in speech applications~\citep{deng2010binary,hinton2012deep,sainath2012auto,gehring2013extracting,zeiler+etal_icassp13, badino2014auto,kamper2015unsupervised}.

We use a stacked AE architecture, as illustrated in Figure~\ref{fig:ae_arch}.
The input layer has 39 units which correspond to the 39-dimensional MFCC or BNFs described in Sections~\ref{sec:mfccs} and \ref{sec:bnf}, respectively.
The output layer is a linear layer that produces the predicted input feature vector $\hat{\vec{x}}$.
The intermediate feature extraction layer has 39 units so that its output corresponds with the dimensionalities of the other feature types.
Implementation and training details are discussed in Section~\ref{sec:exp_setup_feats}.

\subsection{Correspondence autoencoder features}
\label{sec:cae_features}

\begin{figure}[!h]
\centering
\includegraphics[width=0.65\textwidth]{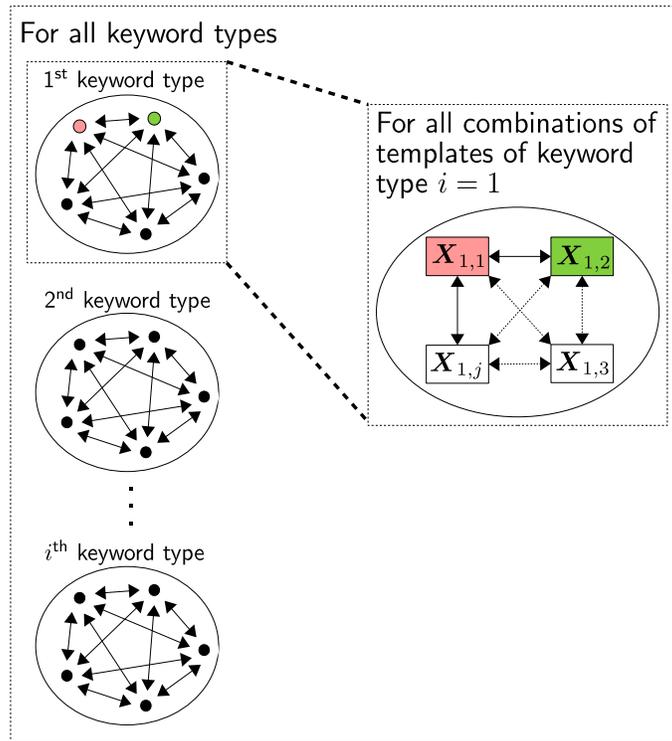}
\caption{
All possible pairwise combinations between keyword templates for each keyword type are used to train the CAE.
$\vec{X}_{i,j}$ is the feature vector sequence of the $j^\textrm{th}$ template of the $i^\textrm{th}$ keyword type.
}
\label{fig:cae_kw_combos}
\end{figure}

Whereas an AE is trained using the same speech feature vectors as input and output, a correspondence autoencoder (CAE) uses feature vectors from different instances of the same keyword type as input and output.
Using the set of isolated keywords (described in Section~\ref{sec:keyword_data} and in Table~\ref{tab:kw_dataset}), we consider all possible pairs of words of the same type, as illustrated in Figure~\ref{fig:cae_kw_combos}.

\begin{figure}[!t]
\centering
\includegraphics[width=0.987\textwidth]{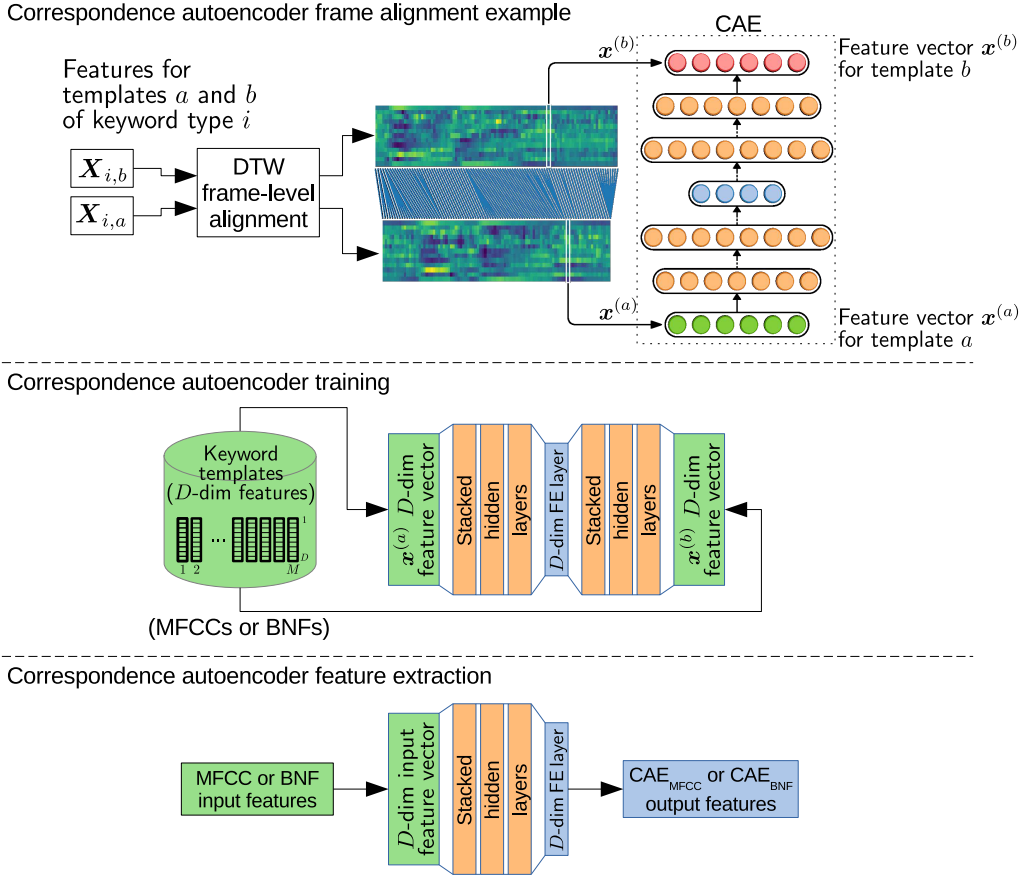}
\caption{
\textit{Top:} Correspondence autoencoder (CAE) training.
The CAE is trained to reconstruct a feature vector extracted from one template of a keyword from an aligned feature vector from another template of the same keyword type.
\textit{Middle:} The CAE is initialised using the pretrained AE weights and fine-tuned on the corpus of isolated keywords recordings.
\textit{Bottom:} After training, the first half of the network (up to the feature extraction (FE) layer) is used as feature extractor.
}
\label{fig:cae_train_extract}
\end{figure}

For each pair, the minimum-cost frame-level alignment between the two words is found by DTW, as shown at the top of Figure~\ref{fig:cae_train_extract}.
Aligned frames are then used as input-output feature vector pairs to train the CAE.
The CAE is therefore trained on pairs of speech features $(\vec{x}^{(a)}, \vec{x}^{(b)})$, where $\vec{x}^{(a)}$ is a feature vector from one keyword, and $\vec{x}^{(b)}$ an aligned feature vector from another keyword of the same type.
If the output of the network in response to $\vec{x}^{(a)}$ is $\hat{\vec{x}}$, then the CAE is trained to minimise the loss $||\hat{\vec{x}} - \vec{x}^{(b)}||^2$.

To obtain useful features, it is essential to pretrain the CAE as a conventional AE~\citep{kamper2015unsupervised}.
Our CAE has the same structure as the AE described in Section~\ref{sec:ae_features} and pretraining follows the same procedure.
After training the AE network, fine-tuning is performed using the set of isolated keywords and the loss described above to obtain the CAE.
In this way, the CAE takes advantage of a large amount of untranscribed data (for initialisation as an AE) as well as a small amount of labelled data (for subsequent fine-tuning as a CAE).
Output features are again extracted from the intermediate 39-dimensional feature extraction layer, as illustrated in the bottom part of Figure~\ref{fig:cae_train_extract}.
Implementation-specific details are given in Section~\ref{sec:exp_setup_feats}.

By training on different templates of the same keyword type, the CAE learns features that are insensitive to factors not common to keyword pairs, such as speaker, gender and channel, while remaining dependent on factors that are, such as the word identity.
Furthermore, the number of input-output pairs on which the CAE is fine-tuned is much larger than the total number of feature vectors in the keyword segments themselves, because all pairwise combinations of different templates of a keyword type are considered.
For example, for the SABN dataset, the keywords contain approximately 120k feature vectors in total, while the pairwise combinations yield approximately two million unique aligned feature vector pairs.
By presenting feature vector pairs to the CAE in both input-output directions, the number of training instances is further doubled to four million.
This represents an advantage over, for example, the training of a CNN keyword spotter, as described in Section~\ref{sec:cnn_classifier}.
 \section{Experimental setup}

\begin{figure}[!ht]
\centering
\includegraphics[width=0.8\textwidth]{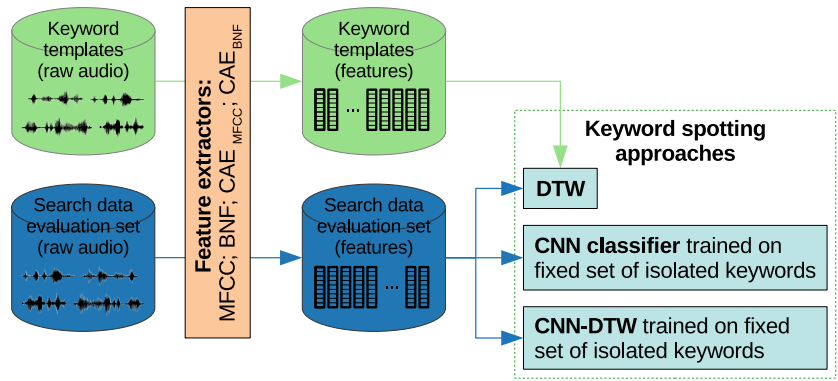}
\caption{Experimental setup for evaluating the different features (MFCCs, BNF, CAE$_{\rm MFCC}$, CAE$_{\rm BNF}$) in combination with the different keyword spotting approaches (DTW, CNN classifier, CNN-DTW).}
\label{fig:systems_eval}
\end{figure}

All experiments are performed independently for both the English and the Luganda data (Section~\ref{sec:data}) but the same experimental procedure is followed for each.
The three keyword spotting approaches covered in Section~\ref{sec:kws_approaches} are used in combination with the six feature types discussed in Section~\ref{sec:features}.
This experimental setup is illustrated in Figure~\ref{fig:systems_eval}.

The keyword spotting approaches and feature extractors are treated as interchangeable modular units.
It is, however, required that the feature type used during training matches the feature type presented at test time.
Table~\ref{tab:kw_feat_combos} shows the combinations of keyword spotting and feature types that we consider.
Our experiments indicate that the AE-based features do not afford any performance benefit over the MFCC or BNF features for the DTW classifier.
In preliminary experiments we found that the same is true for the CNN and CNN-DTW classifiers.
Therefore, CNN and CNN-DTW results will not be presented for autoencoder features.
All neural networks are implemented using the Theano 1.0.2~\citep{al2016theano} and Lasagne 0.2.dev1~\citep{lasagne} toolkits.

\begin{table}[!hb]
\footnotesize
\caption{The combinations of keyword spotting approaches (1\textsuperscript{st} column) evaluated against the feature types (1\textsuperscript{st} row).
}\label{tab:kw_feat_combos}
\renewcommand{\arraystretch}{1.1}
\begin{tabular*}{\textwidth}{@{\extracolsep{\fill}} lcccccc @{} }
\toprule
Architecture  & MFCC & BNF & AE$_{\rm MFCC}$ & AE$_{\rm BNF}$ & CAE$_{\rm MFCC}$ & CAE$_{\rm BNF}$ \tabularnewline
\midrule
CNN classifier  & \checkmark & \checkmark & --         & --         & \checkmark & \checkmark \tabularnewline
DTW             & \checkmark & \checkmark & \checkmark & \checkmark & \checkmark & \checkmark \tabularnewline
CNN-DTW         & \checkmark & \checkmark & --         & --         & \checkmark & \checkmark \tabularnewline
\bottomrule
\end{tabular*}
\end{table}

\subsection{Keyword spotting approaches}
\label{sec:exp_setup_kws}

The CNN and DTW keyword classifiers described in Section~\ref{sec:kws_approaches} are used as baselines for comparative evaluation with our proposed CNN-DTW approach.

\subsubsection{CNN keyword classifier}
\label{sec:exp_setup_cnn_classifier}

As our first baseline, we consider the direct application of a CNN classifier as described in Section~\ref{sec:cnn_classifier}.
We perform supervised training on the network using the recorded isolated keywords described in Table~\ref{tab:kw_dataset}.

The dimensionality of the input layer is 60$\times$39.
Since the isolated keyword utterances vary in duration, we resample the time dimension using cubic interpolation to obtain a feature matrix with these dimensions.

The input layer is followed by three convolutional layers with 64 11$\times$11, 128 7$\times$7 and 256 5$\times$5 filters, respectively.
Filter strides are set to one.
The second and third convolutional layers are each followed by a max pooling layer and the last convolutional layer by three dense layers with 500, 100 and 300 units, respectively.
The first and the third dense layers are each followed by a dropout layer with probability $p=0.5$.
The network is terminated by a dense softmax layer with as many units as there are keyword types in the respective datasets (Table~\ref{tab:kw_dataset}).
ReLUs are used as activation functions in all other layers.
Nesterov momentum is used to optimise the categorical cross entropy loss given in Equation~\eqref{eq:multiclass}, with a learning rate which is adjusted linearly from $10^{-4}$ to $10^{-6}$.
The architectural choices have been optimised for performance on a held-out portion of the English isolated keywords set (Table~\ref{tab:kw_dataset}).
The network is trained for 1000 epochs, using early stopping on the same held-out set.

During initial experimentation, we found that the inclusion of a rejection class trained on negative training examples always led to a deterioration in performance, and that using only positive training examples led to the best performing system.
Therefore, no rejection class has been included.

\subsubsection{DTW keyword classifier}
For the second baseline system, DTW (Section~\ref{sec:dtw_classifier}) is performed directly on the evaluation utterances for each exemplar of a keyword.
The best similarity score among all exemplars of a keyword is used to make a classification decision.
This system serves as our topline, since the same DTW scores are used as targets during supervised training of the CNN-DTW keyword classifier.
It is also by far the most computationally expensive of the approaches we consider.
Consequently, to ensure reasonable runtimes, experiments were performed on a high performance computing cluster.

\subsubsection{CNN-DTW keyword classifier}
\label{sec:exp_setup_cnn_dtw}

For the CNN-DTW keyword classification approach that we propose, the training of the CNN is supervised by DTW similarity scores, as described in Section~\ref{sec:cnn_dtw_classifier}.
Here the CNN consists of ten convolutional layers, the first of which consists of 80 39$\times$10 filters.
The following three convolutional layers consist of 80 1$\times$10 filters, followed by three layers with 256 1$\times$10 filters and three layers with 512 1$\times$10 filters.
All filter strides are set to one.
The last convolutional layer is followed by a global max pooling layer, two 3000-unit dense layers, each with a dropout probability of $p=0.5$, and a learning rate changing linearly from $10^{-4}$ to $10^{-5}$ used with Adam optimisation \citep{Kingma2014}.
Leaky ReLU activations with $\alpha = 1/3$ are used for all hidden layers.
The final output layer is dense with sigmoid activation functions and as many units as keyword types in the respective datasets (Table~\ref{tab:kw_dataset}).
The architectural choices have been optimised for performance using the automatically obtained English development set DTW scores (Table~\ref{tab:radio_data}).
The summed cross-entropy loss in Equation~\eqref{eq:summed_binary} is optimised by training the network for 1000 epochs with early stopping.
The loss in Equation~\eqref{eq:summed_binary} is computed using the automatically obtained development set DTW scores.
Hence, no transcriptions are used during either training or validation.

\subsection{Features and feature extractors}
\label{sec:exp_setup_feats}
The feature extractors described in Section~\ref{sec:features} are applied to the radio broadcast data described in Table~\ref{tab:radio_data} and the isolated keyword template data in Table~\ref{tab:kw_dataset} to produce the MFCC, BNF, AE$_{\rm MFCC}$, AE$_{\rm BNF}$, CAE$_{\rm MFCC}$ and CAE$_{\rm BNF}$ features that are used in the experiments described in Section~\ref{sec:exp_results}.

\subsubsection{MFCC features}
\label{sec:mfcc_setup}
Using the HTK toolkit~\citep{young2002htk}, we extract 13 MFCCs from the speech audio and append velocity and acceleration coefficients, resulting in the 39-dimensional MFCC feature vectors described in Section~\ref{sec:mfccs}.
This is followed by per-utterance cepstral mean and variance normalisation.

\subsubsection{Bottleneck (BNF) features}
\label{sec:bnf_setup}
For the bottleneck feature extractor introduced in Section~\ref{sec:bnf}, we use a six-layer time-delay neural network (TDNN) \citep{peddinti2015time} trained on ten languages from the GlobalPhone corpus, as described in~\citep{Hermann2018}.
For speaker adaptation, a 100-dimensional i-vector is appended to the 40-dimensional high resolution MFCC input features.
The TDNN is trained using a block-softmax \citep{Grezl2014}, with hidden layers shared between all languages and with ten separate output layers, one for each language.
Each of the six hidden layers has 625 dimensions and is followed by a 39-dimensional bottleneck layer with ReLU activations and batch normalisation.
Each language then has its own 625-dimensional affine transformation and softmax layer.
Training is performed for two epochs with stochastic gradient descent and closely follows the Babel multilingual recipe \citep{Trmal2017} of the Kaldi ASR toolkit~\citep{Povey2011}.
The training data consists of 198 hours of multilingual GlobalPhone data, as described in Section~\ref{sec:bnf_data}.

\subsubsection{Autoencoder (AE) features}
\label{sec:ae_setup}
We use a stacked AE architecture, as introduced in Section~\ref{sec:ae_features}.
The input layer of the network has 39 units which correspond to the 39-dimensional MFCCs or BNFs described in Sections~\ref{sec:mfcc_setup} and \ref{sec:bnf_setup}, respectively.
We use a symmetrical stacked architecture: the input layer is followed by eight 100-unit hidden layers, a 39-dimensional feature extraction layer, another eight 100-unit hidden layers, and a final 39-dimensional output layer.
The weights of the hidden layers before the feature extraction layer are tied to the weights of the corresponding hidden layers after the feature extraction layer.
Hence, the weights of the last hidden layer are the transpose of the weights of the first hidden layer, the weights of the second-to-last hidden layer are the transpose of those of the second hidden layer, and so forth.
We use tanh activations throughout.
These architectural choices were optimised on development data.
The network is trained on the radio broadcast training sets described in Table~\ref{tab:radio_data}, disregarding the transcriptions.
Two feature types, MFCCs and BNFs, are considered as features to train two separate AE extractors, denoted by AE$_{\rm MFCC}$ and AE$_{\rm BNF}$, respectively.
During training, an MFCC or BNF feature vector is presented at the input layer of the network, while the same vector is used as the training target.
We use the ADADELTA optimisation algorithm \citep{Zeiler2012}, minimising the squared error loss.
Layer-wise pretraining,  as described in Section~\ref{sec:ae_features}, is first performed, with each individual layer trained for five epochs.
The entire network is then fine-tuned for a further five epochs.
A batch size of 2048 feature vectors is used throughout.
The 39-dimensional hidden layer becomes the feature extraction layer when the AE is used as a feature extractor and provides the features used in the AE$_{\rm MFCC}$ and AE$_{\rm BNF}$ experiments.

\subsubsection{Correspondence autoencoder (CAE) features}
\label{sec:cae_setup}

The CAE, introduced in Section~\ref{sec:cae_features}, uses the same neural network architecture used by the AE described in the previous section.
In fact, the AE itself is used to initialise the training of the CAE.
The CAE is distinguished from the AE in that it is further fine-tuned on the features extracted from the isolated keyword templates (Table~\ref{tab:kw_dataset}) employing the process and reconstruction-like loss described in Section~\ref{sec:cae_features}.
As for the AE, two feature types, MFCCs and BNFs, are used to train two separate CAE feature extractors, denoted by CAE$_{\rm MFCC}$ and CAE$_{\rm BNF}$ respectively.
The ADADELTA optimisation algorithm \citep{Zeiler2012} was again employed, training the network for 120 epochs using a batch size of 2048 feature vectors.
The 39-dimensional feature extraction layer provides the features used in the CAE$_{\rm MFCC}$ and CAE$_{\rm BNF}$ experiments.

\subsection{Evaluation metrics}

Keyword spotting performance is assessed using a number of standard metrics.
The receiver operating characteristic (ROC) is obtained by plotting the false positive rate against the true positive rate as the keyword detection threshold is varied.
The area under this curve (AUC) is used as a single metric that characterises the classifier across all operating points.
The equal error rate (EER) is the point at which the false positive rate equals the false negative rate.

Precision at 10 ($P$@10) and precision at $N$ ($P$@$N$) are the proportion of correct keyword detections among the top 10 and top $N$ hits, respectively, with $N$ corresponding to the number of true occurrences of a keyword type in the evaluation data.
For example, for a single keyword type, $P@10 = {}^2/_5$ \newline indicates that among the ten top-ranked utterances, four utterances contained true occurrences of the keyword, while six were false retrievals.
The reported $P$@10 and $P$@$N$ values are the averages calculated over all keyword types.
For example, for $P$@10:

\begin{equation}
\mbox{average }P@10 = \frac{1}{K} \sum_{k=1}^K (P@10)_k
\end{equation}
where $K$ is the number of keyword types. The same applies to the average~$P$@$N$, where $P$@10 is substituted for $P$@$N$.
 \section{Experiments}
\label{sec:exp_results}

Table~\ref{tab:results_dev_h} shows the performance on the English development data of the various keyword spotting approaches trained on the different feature types.
Figures~\ref{fig:rocs_sabn_dev_h} and \ref{fig:rocs_sabn_luganda_kws_h} show the corresponding ROC curves for the various systems on development data.
Based on these development results, we considered only the best feature types for each keyword spotting approach for final evaluation on the English and Luganda test data.
These results are given in Table~\ref{tab:results_test_h}.

\begin{table}[!hb]
\small
\caption{English development set results. Values are expressed as percentages (\%), hence an AUC of 100\% represents an area under the curve of one. The best feature type for each keyword spotting approach is set in bold.}\label{tab:results_dev_h}
\renewcommand{\arraystretch}{1.1}
\begin{tabular*}{\textwidth}{@{\extracolsep{\fill}} l rrrr @{} }
\toprule

\multicolumn{5}{c}{{\bf{CNN keyword classifier}}} \\
\midrule
Features       & AUC & EER & \pten & \pN
\tabularnewline
\midrule
MFCC             & 55.35 & 46.43 &  6.00 &  4.13 \tabularnewline
CAE$_{\rm MFCC}$ & 71.90 & 34.02 &  8.00 &  7.58 \tabularnewline
BNF              & 77.14 & 29.83 & 16.50 & 13.66 \tabularnewline
CAE$_{\rm BNF}$  & \textbf{77.98} & \textbf{29.09} & \textbf{20.05} & \textbf{16.75} \tabularnewline
\midrule

\multicolumn{5}{c}{{\bf{DTW keyword classifier}}} \\
\midrule
Features       & AUC & EER & \pten & \pN
\tabularnewline
\midrule
MFCC             & 70.59 & 35.25 & 15.50 &  9.40 \tabularnewline
AE$_{\rm MFCC}$  & 69.22 & 36.99 & 16.00 &  9.02 \tabularnewline
CAE$_{\rm MFCC}$ & 73.68 & 33.10 & 24.00 & 14.27 \tabularnewline
BNF              & 74.60 & 32.63 & 17.50 & 14.08 \tabularnewline
AE$_{\rm BNF}$   & 75.58 & 32.03 & 17.50 & 13.80 \tabularnewline
CAE$_{\rm BNF}$  & \textbf{83.93} & \textbf{23.68} & \textbf{41.25} & \textbf{31.66} \tabularnewline
\midrule

\multicolumn{5}{c}{{\bf{CNN-DTW keyword classifier}}} \\
\midrule
Features       & AUC & EER & \pten & \pN
\tabularnewline
\midrule
MFCC             & 61.85 & 41.09 &  2.00 &  2.81 \tabularnewline
CAE$_{\rm MFCC}$ & 69.40 & 36.61 & 14.75 & 10.83 \tabularnewline
BNF              & 73.56 & 32.71 & 14.00 &  9.96 \tabularnewline
CAE$_{\rm BNF}$  & \textbf{83.28} & \textbf{24.16} & \textbf{38.00} & \textbf{28.28} \tabularnewline
\bottomrule
\end{tabular*}
\end{table}

\begin{figure}[!ht]
\footnotesize
\centering
\renewcommand{\arraystretch}{1.7}
\begin{tabular}{@{} c  c @{}}
\includegraphics[width=0.48\textwidth]{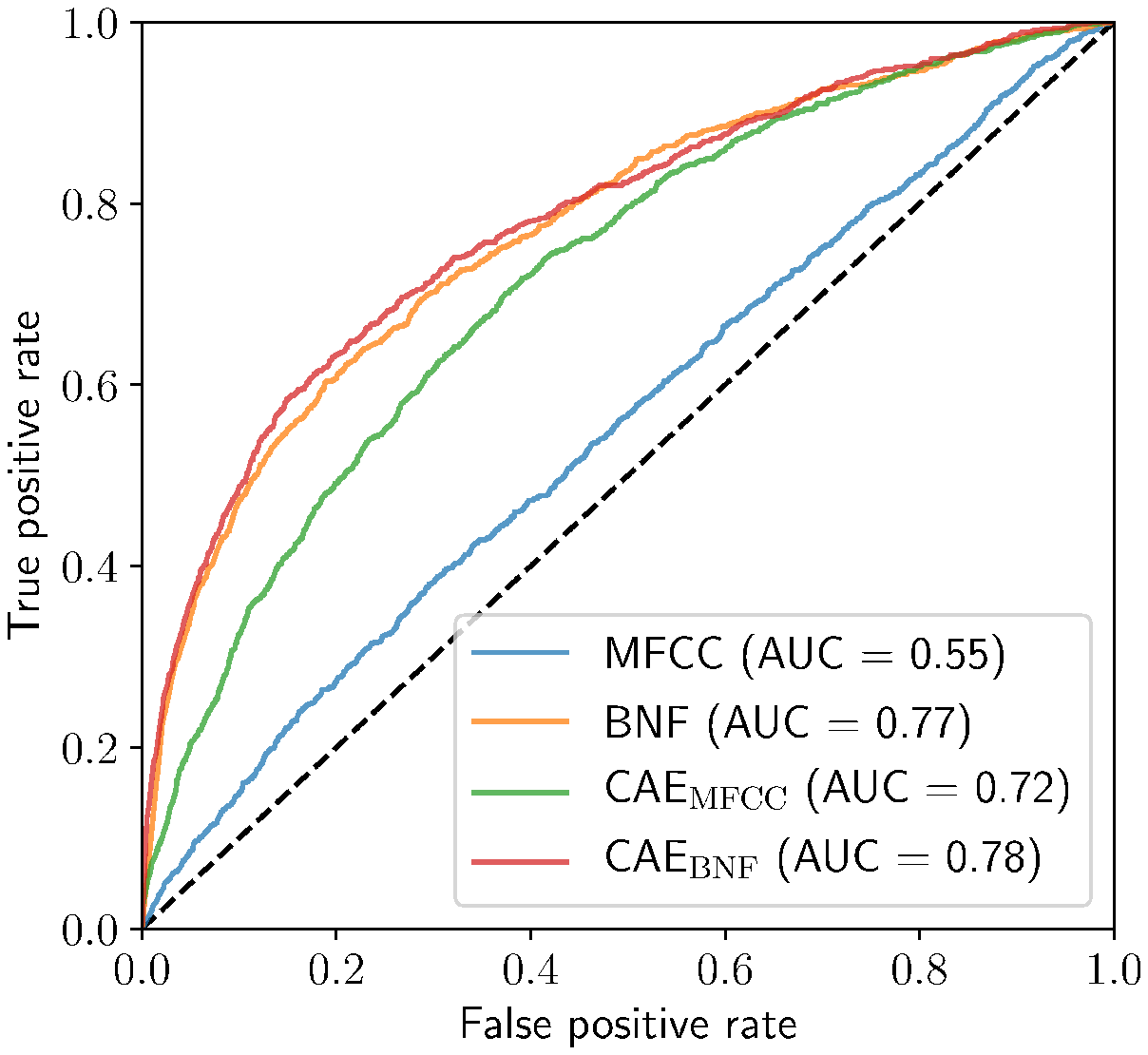}  &  \includegraphics[width=0.48\textwidth]{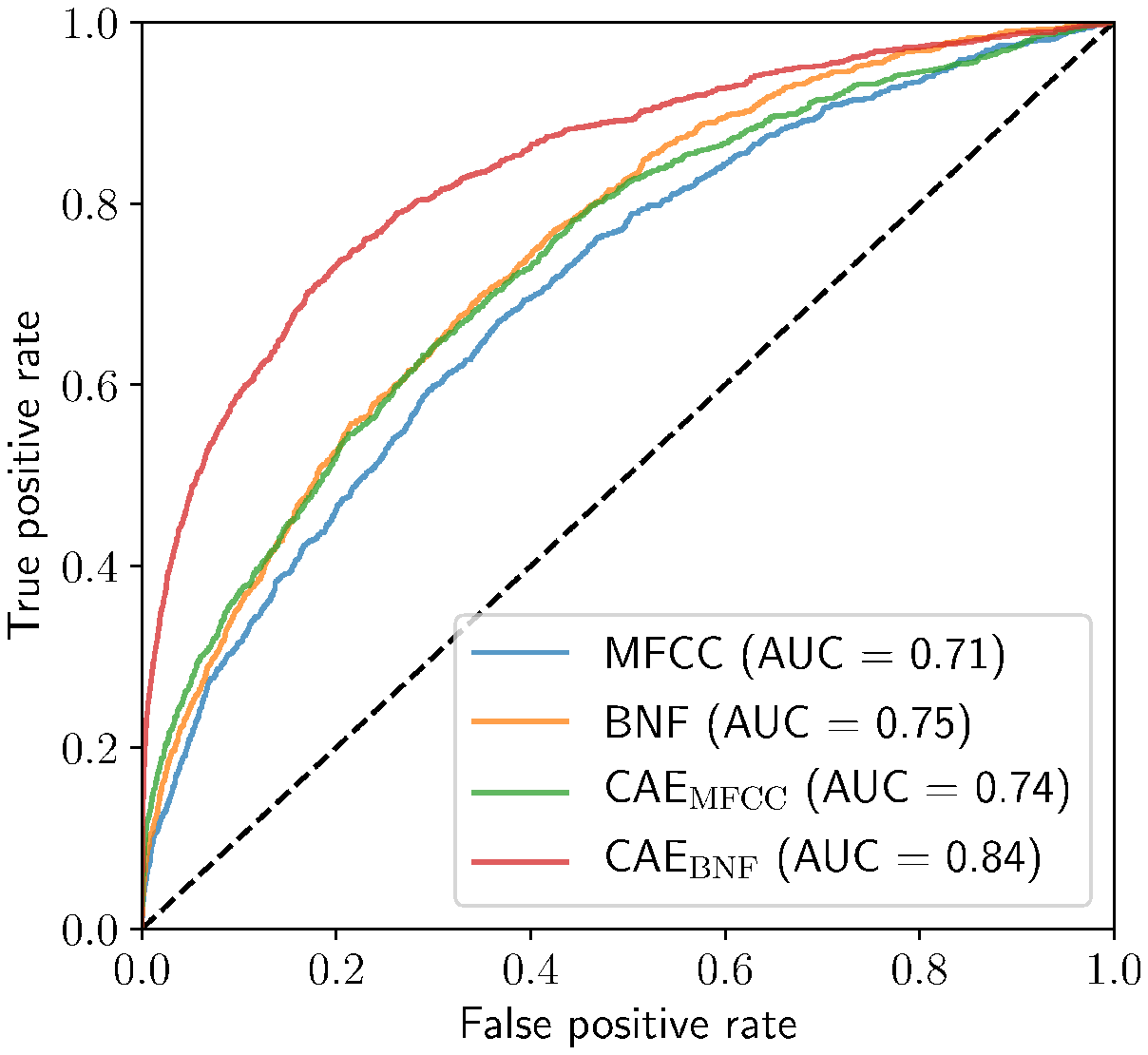}  \tabularnewline
(a) CNN classifier & (b) DTW classifier \tabularnewline
\multicolumn{2}{c}{\includegraphics[width=0.48\textwidth]{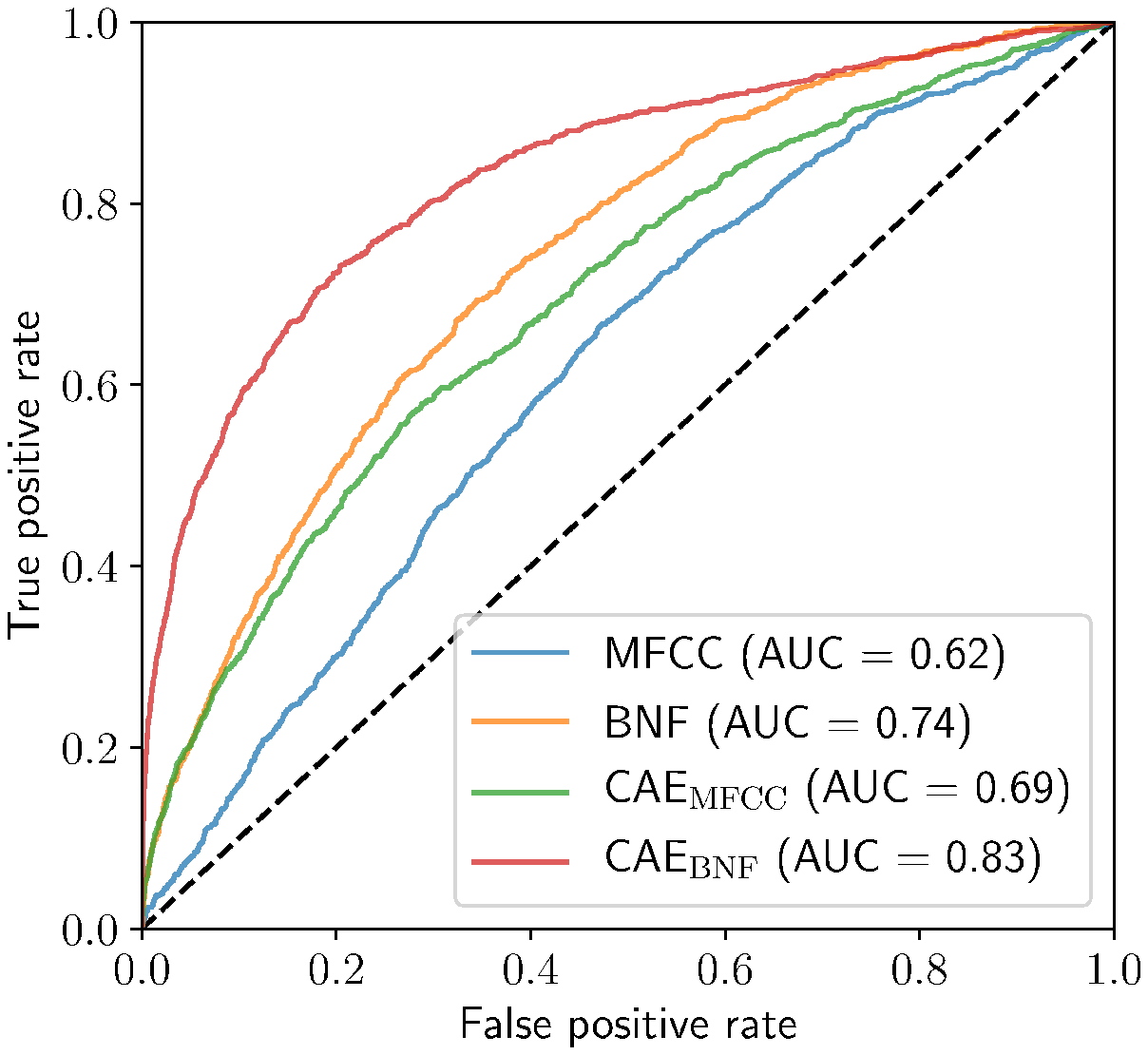}} \tabularnewline
\multicolumn{2}{c}{(c) CNN-DTW} \tabularnewline
\end{tabular}
\caption{ROC curves obtained on the English development set for the (a) CNN, (b) DTW and (c) CNN-DTW keyword spotting approaches using the different feature types.}
\label{fig:rocs_sabn_dev_h}
\end{figure}

\begin{table}[!h]
\small
\caption{English and Luganda test set results. Values are expressed as percentages (\%), i.e.\ an AUC of 100\% represents an area under the curve of one. The best feature type for each keyword spotting approach is shown in bold.
}\label{tab:results_test_h}
\renewcommand{\arraystretch}{1.1}
\begin{tabular*}{\textwidth}{@{\extracolsep{\fill}} l rrrr rrrr @{} }
\toprule
 & \multicolumn{4}{c}{English} & \multicolumn{4}{c}{Luganda} \tabularnewline
\cmidrule(lr){2-5} \cmidrule(l){6-9}
Features & AUC & EER & \pten & \pN & AUC & EER & \pten & \pN \tabularnewline
\midrule
\multicolumn{9}{c}{\bf{CNN keyword classifier}} \tabularnewline
\midrule
CAE$_{\rm MFCC}$ & 70.45 & 34.85 &  7.50 &  5.48 & 44.02 & 54.94 &  3.33 &  2.84 \tabularnewline
BNF              & \textbf{76.55} & \textbf{30.62} &  9.75 & 10.91 & \textbf{51.07} & \textbf{49.18} &  \textbf{1.11} &  \textbf{2.10} \tabularnewline
CAE$_{\rm BNF}$  & 75.32 & 32.06 & \textbf{15.00} & \textbf{13.01} & 43.38 & 53.07 &  \textbf{1.11} &  1.96 \tabularnewline
\midrule
\multicolumn{9}{c}{\bf{DTW keyword classifier}} \tabularnewline
\midrule
MFCC             & 72.38 & 33.42 & 15.25 &  8.83 & 71.26 & 35.32 & 16.67 & 12.89 \tabularnewline
BNF              & 75.06 & 32.22 & 20.00 & 12.85 & \textbf{77.71} & \textbf{29.47} & 22.78 & 16.59 \tabularnewline
CAE$_{\rm BNF}$  & \textbf{84.55} & \textbf{23.61} & \textbf{42.75} & \textbf{30.21} & 75.96 & 32.04 & \textbf{37.22} & \textbf{30.34} \tabularnewline
\midrule
\multicolumn{9}{c}{\bf{CNN-DTW keyword classifier}} \tabularnewline
\midrule
CAE$_{\rm MFCC}$ & 70.44 & 34.56 & 17.75 & 10.71 & 69.33 & 37.58 & 16.67 &  9.21 \tabularnewline
BNF              & 73.84 & 33.00 & 20.50 &  9.90 & \textbf{76.96} & \textbf{29.85} & 13.89 & 10.68 \tabularnewline
CAE$_{\rm BNF}$  & \textbf{83.31} & \textbf{24.60} & \textbf{41.25} & \textbf{27.97} & 74.85 & 32.85 & \textbf{33.33} & \textbf{22.60} \tabularnewline
\bottomrule
\end{tabular*}
\end{table}

\subsection{Features and feature extractors}
\label{sec:experiments_features}

Before evaluating the different keyword spotting approaches, we consider the performance that is achieved using the different feature types.
Firstly, we compare CAE$_{\rm MFCC}$ features to standard BNFs.
These features respectively use a small amount of labelled in-domain data and a large amount of labelled data from several out-of-domain languages.
We observe in Table~\ref{tab:results_dev_h} that they either perform similarly or that the BNFs perform slightly better.
Specifically, in the DTW system the features perform very similarly, as was also found in~\citep{menon+etal_interspeech19}.
However, for the CNN as well as the CNN-DTW system (Table~\ref{tab:results_dev_h} and Figure~\ref{fig:rocs_sabn_dev_h}), BNFs perform slightly better.
When we consider the English and Luganda test results for the CNN and CNN-DTW classifiers in Table~\ref{tab:results_test_h}, we also see that BNFs generally
outperform the CAE$_{\rm MFCC}$ features, although there are some metrics on which the latter performs better.
Overall, the results {show} that both standard BNFs and CAE$_{\rm MFCC}$ features provide improvements over traditional MFCC features.
This suggests that both a small amount of labelled in-domain data (CAE$_{\rm MFCC}$) and a large amount of labelled data from several out-of-domain languages (BNFs) are beneficial for feature learning.

The English development results in Table~\ref{tab:results_dev_h} also show that the CAE$_{\rm MFCC}$ features outperform the MFCC baseline as well as the AE$_{\rm MFCC}$ features for all three considered system architectures and all four performance measures, with relative improvements in AUC of around 5\%.
Similarly, the same table shows that CAE$_{\rm BNF}$ features outperform BNF and AE$_{\rm BNF}$ feature types.
We therefore see that CAE-based features provide consistent improvements over the AE-based counterparts.

Our best overall results on the English development data in Table~\ref{tab:results_dev_h}  are achieved using the CAE$_{\rm BNF}$ features.
The improvements are consistent over all three keyword spotting approaches and all four metrics considered, and are reflected by the ROC curves in Figure~\ref{fig:rocs_sabn_dev_h}.
These features achieve precision values that are at least 1.65 times better than the closest competitor.
Specifically, for the topline DTW system, the AUC and EER are improved by 12.5\% and 27.4\%, respectively, when using CAE$_{\rm BNF}$ features instead of standard BNFs, and achieve more than double the precision when considering the top-10 and top-$N$ retrievals.
When compared to the baseline MFCCs, these AUC and EER improvements increase to 18.9\% and 32.8\%, respectively.

When considering the test set results in Table~\ref{tab:results_test_h}, there are some cases where BNFs perform slightly better than  CAE$_{\rm BNF}$ features for the CNN and the CNN-DTW approaches.
For the CNN classifier, BNFs achieve better AUC and EER than the CAE$_{\rm BNF}$ features for both English and Luganda.
However, we also see that the CNN classifier performs very poorly for Luganda in terms of \pten and $P@N$, and this makes it difficult to draw conclusions.
For the CNN-DTW system, {the} BNFs still lead to slightly better AUC and EER figures on the Luganda test set, but in this case substantially better AUC and EER are seen for English.
Moreover, \pten and \pN are substantially better for the CAE$_{\rm BNF}$ features, achieving more than twice as many correct retrievals for both English and Luganda than BNFs.
This is despite the Luganda systems using hyperparameters optimised only on the English development set in order to reflect the practical resource-constrained application of the keyword spotter.
Considering the development and test results together across both languages, the experiments show that CAE$_{\rm BNF}$ features are able to robustly combine the benefits of learning from well-resourced non-target language data with the benefits of fine-tuning on a small amount of labelled in-domain speech.

\subsection{Keyword spotting approaches}

We next turn to a comparison of the different keyword spotting approaches.
Considering both the English development results in Table~\ref{tab:results_dev_h} and the test results in Table~\ref{tab:results_test_h}, the DTW system consistently outperforms both the CNN and the CNN-DTW keyword classifiers for the same feature type.
In most cases the baseline CNN classifier yields the poorest performance of the three approaches.
Although the AUC and EER of the CNN and the CNN-DTW classifiers are comparable for English when using  BNF and CAE$_{\rm MFCC}$ features, performance is far worse when regarding the precision metrics.
This trend is however not observed for the Luganda test set, where the CNN-DTW always outperforms the CNN classifier.

\begin{figure}[!ht]
\footnotesize
\centering
\begin{tabular}{@{} c c @{}}
\includegraphics[width=0.5\textwidth]{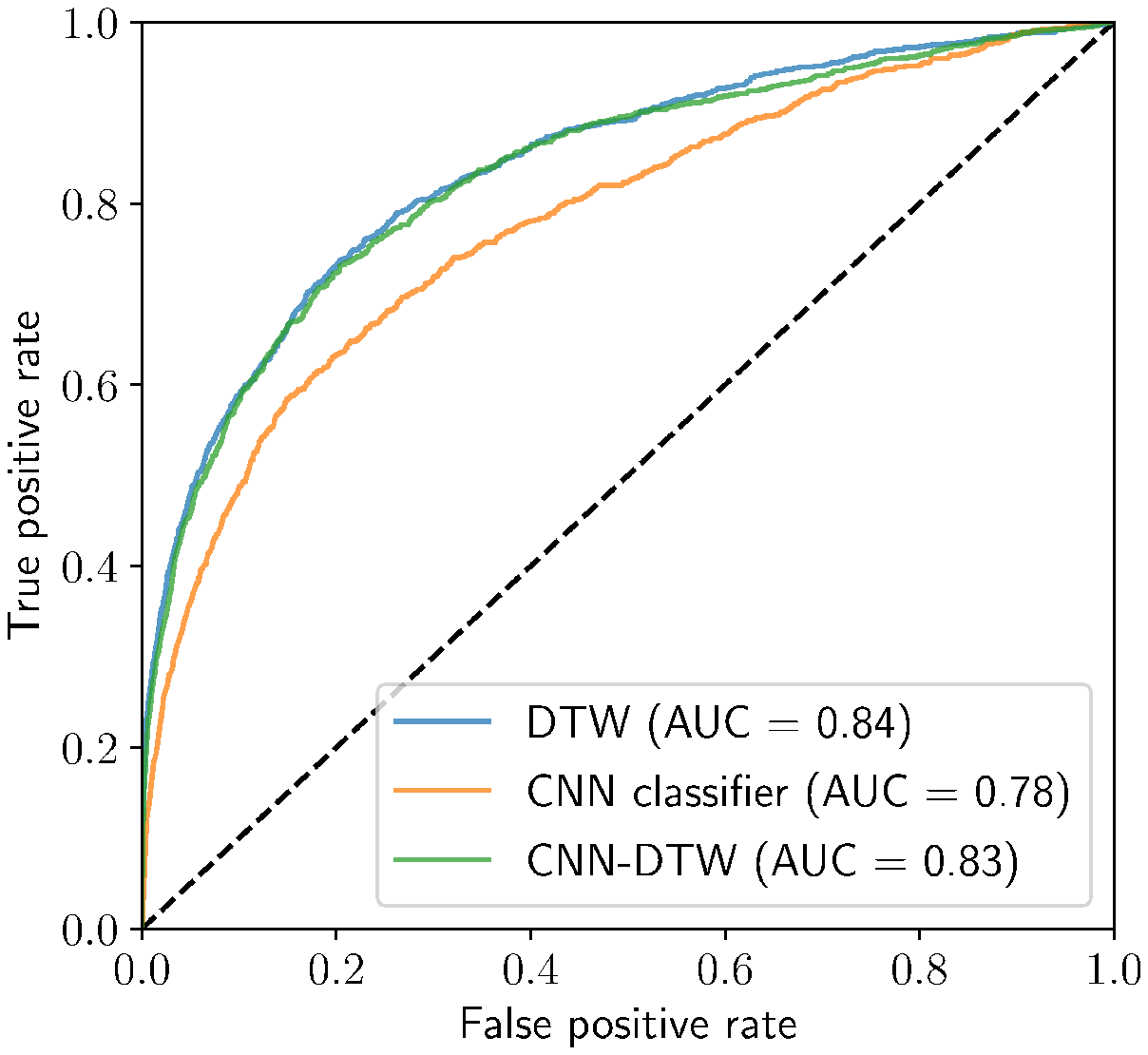} & \includegraphics[width=0.5\textwidth]{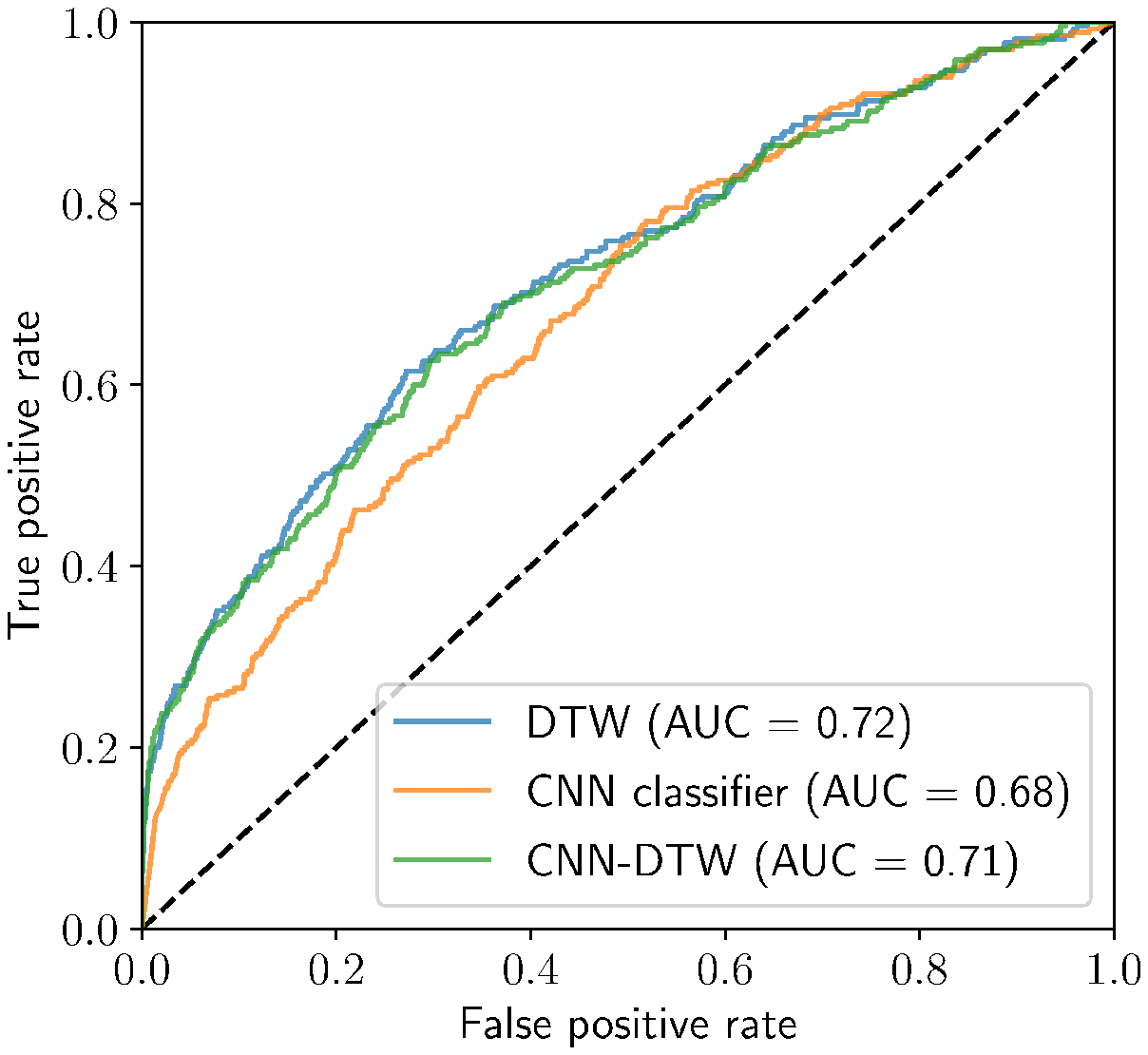} \tabularnewline
(a) English & (b) Luganda \tabularnewline
\end{tabular}
\caption{ROC curves for the different keyword spotting approaches using the CAE$_{\rm BNF}$ features on the (a) English and (b) Luganda development data.}
\label{fig:rocs_sabn_luganda_kws_h}
\end{figure}

The DTW system provides the targets with which the CNN-DTW system is trained, and hence represents an upper bound on the performance we can expect from the CNN-DTW system.
The difference in performance between the DTW and the CNN-DTW systems is smallest when using the BNF and CAE$_{\rm BNF}$ features---in this case the two systems have nearly the same performance.
This is also evident when considering the ROC curves on the English and Luganda data shown in Figure~\ref{fig:rocs_sabn_luganda_kws_h}.

\begin{table}
\footnotesize
\caption{
Computational complexity analysis.
The number of multiplications is calculated assuming a 15-second input audio segment.
The runtimes are average measurements, also for a 15-second input audio segment.
The CPU experiments were executed on a single core of an Intel Core i7-8700K.
The GPU experiments were executed as a single process using one GeForce GTX 1080 Ti.
}
\label{tab:results_runtimes_h}
\renewcommand{\arraystretch}{1.1}
\begin{tabular*}{\textwidth}{@{\extracolsep{\fill}} lrrr @{}}
\toprule
Approach                   &             DTW  &         CNN  &    CNN-DTW  \tabularnewline
\midrule
Multiplications  &  $9.54\times10^{12}$  & $146.29\times10^{9}$ & $88.06\times10^{6}$ \tabularnewline
\midrule
CPU (Intel)                &   227 s &   82.50 s\phantom{m} & 164.74 ms \tabularnewline
\midrule
GPU                        &    N/A &  925.26 ms & 3.40 ms \tabularnewline
\bottomrule
\end{tabular*}
\end{table}

The advantage of the proposed CNN-DTW approach lies in its extremely fast execution.
Table~\ref{tab:results_runtimes_h} shows a comparison of the computational complexity in terms of the number of multiplications and also the execution times on CPU and GPU platforms.
The number of multiplications was calculated assuming an input audio segment with a fixed length of 15 seconds.
For DTW, 1160 keyword templates and a frame step of three were used.
For the CNN classifier the frame step is one.
These values reflect our test setting as well as a feasible operational configuration.
According to these calculations, one application of the CNN classifier requires $101\times10^6$ multiplications, which is the same order of magnitude as the CNN-DTW.
However, the CNN classifier has to be applied several times as it is swept across the length of an utterance.
For a 15-second utterance, this increases the total number of multiplication by roughly a factor 1500 to $146.29\times10^{9}$.
The CNN classifier therefore requires $\approx$\,65~times fewer multiplications than DTW, while CNN-DTW requires $\approx$\,1661~times fewer multiplications than the CNN classifier and $\approx$\,$10^5$~times fewer multiplications than DTW.

Although the number of multiplications provides an indication of the complexity of each approach, it is not an exact indicator of physical runtime.
We have therefore measured the average CPU and GPU runtimes required to process a 15-second input audio segment on one CPU core and one GPU device, respectively.
An Intel Core i7-8700K is used for the CPU measurement and a GeForce GTX 1080 Ti for the GPU measurement.
Table~\ref{tab:results_runtimes_h} shows that, in terms of CPU execution time, DTW is $\approx$\,2.8~times slower than the CNN classifier and $\approx$\,1378~times slower than CNN-DTW, while
the CNN classifier is $\approx$\,500~times slower than CNN-DTW.

Our target application requires fast processing.
The acceleration afforded by a GPU is therefore desirable.
The two neural network architectures can take advantage of GPU acceleration while DTW cannot.
In the case of GPU accelerated processing, the CNN and CNN-DTW approaches outperform DTW by several orders of magnitude.
Although direct comparison of the CPU and GPU runtimes is perhaps not fair, it reflects the advantage that GPU-based computation, and consequently the CNN-DTW keyword spotting approach, can provide in a practical setting using readily-available consumer hardware.
The efficiency improvements of CNN-DTW over the DTW approach are noteworthy because of the very small performance penalty incurred by choosing the former over the latter when using CAE$_\textrm{BNF}$ features.
 \section{Conclusion}

We have investigated methods to learn features and improve computational efficiency for ASR-free keyword spotting in a severely resource-constrained setting.
Our experimental evaluation considered two languages: South African English and Luganda.
The first is fairly well resourced and therefore allows careful development, while the second represents a realistic under-resourced scenario.
We considered the specific and practically relevant setting where, for system development, we have at our disposal only a small in-domain corpus consisting of isolated spoken keywords, a larger but untranscribed corpus of in-domain speech, and large transcribed out-of-domain corpora in unrelated languages.

For feature learning, we compared baseline mel-frequency cepstral coefficients (MFCCs) and multilingual bottleneck features (BNFs) to autoencoder (AE) and correspondence autoencoder (CAE) features.
The BNF extractor is trained on labelled data from well-resourced out-of-domain languages.
The CAE is first pretrained as the AE on a large but unlabelled in-domain speech corpus, and then fine-tuned on a small set of in-domain labelled keywords.
By using all possible utterance pairings of the same keyword type as input-output pairs during training, the number of training examples available to the CAE is drastically increased.
Our experiments show that the BNF and CAE architectures are complementary, with best overall performance achieved when using the BNFs as input to the CAE.
This allows us to take advantage of all the data sources at our disposal: the large out-of-domain labelled data (for training the BNF extractor), the large in-domain unlabelled data (for pretraining the CAE), and the small set of in-domain labelled data (for fine-tuning the CAE).

The benefits of using the CAE on top of BNF features is shown here directly in an extrinsic keyword spotting task that uses features obtained from a lightly supervised neural network model.
In contrast to the work of~\citep{Hermann2018,hermann+etal_submitted18}, where discovered word pairs were used for unsupervised CAE training and the benefit of CAE training on top of BNFs were inconclusive in an isolated word discrimination task, we obtain consistent improvements in our setting.

A convolutional neural network (CNN) was trained to predict the DTW alignment costs resulting from the matching of the labelled keywords to the unlabelled in-domain data.
The resulting CNN-DTW keyword classifier is much faster than direct DTW, since no alignment is required.
Furthermore, this CNN-DTW system exhibits best performance when using the combination of BNFs and CAE as input features.
It also exhibits better performance than a more conventional CNN keyword classifier trained only on the labelled keywords.
We conclude that the CNN-DTW architecture we propose is a computationally feasible approach to ASR-free keyword spotting that can be applied in situations where the compilation of a traditional annotated speech corpus is not possible.

Our ongoing work includes the application of the approach we have described to several languages of current interest to humanitarian relief monitoring, including Somali, Rutooro and Lugbara.

\section*{Acknowledgements}
Dynamic time warping computations were performed using the Stellenbosch University's High Performance Computing Cluster (Rhasatsha): \url{http://www.sun.ac.za/hpc}.
Herman Kamper was supported by a Google Faculty Award.

\bibliographystyle{elsarticle-harv}
\bibliography{main}

\end{document}